\documentclass{PoS}

\def\EE8{{\rm E_8\times E_8'}}

\newcommand{\gev}{\,\textrm{GeV}}

\newcommand{\eV}{\,\mathrm{eV}}
\newcommand{\Mp}{M_{\rm P}}

\newcommand{\Mg}{{M_{\rm GUT}}}

\newcommand{\Ude}{U(1)$_{\rm de}$}
\newcommand{\UPQ}{U(1)$_{\rm PQ}$}

\newcommand{\Z}{{\bf Z}}
\newcommand{\delq}{{\delta_{\rm CKM}}}
\newcommand{\dell}{{\delta_{\rm PMNS}}}

\def\fourb{\overline{\bf 4}\,}
\def\four{{\bf 4}}

\def\fiveb{\overline{\bf 5}\,}
\def\five{{\bf 5}}
\def\tenb{\overline{\bf 10}\,}
\def\ten{{\bf 10}}
\def\one{{\bf 1}}

\def\Qem{Q_{\rm em}}
\def\GGsu5{{SU(5)$_{\rm GG}$}}
\def\flip{{SU(5)$_{\rm flip}$}}

\newcommand{\ie}{{\it i.e.~}}
\newcommand{\etal}{{\it et al.}\,}
  
\title{Towards unification of GUT families}

\ShortTitle{Unification of GUT families}

\author{\speaker{Jihn E. Kim}
\thanks{Supported by the NRF grant(No. 2005-0093841) funded by the Korean Government.}\\
  Department of Physics and Astronomy, Seoul National University, 1 Gwanakro, Gwanak-Gu, Seoul 08826, Republic of  Korea, and \\
       Department of Physics, Kyung Hee University, 26 Gyungheedaero, Dongdaemun-Gu, Seoul 02447, Republic of Korea, and \\ 
Center for Axion and Precision Physics Research (IBS),
  291 Daehakro, Yuseong-Gu, Daejeon 34141, Republic of Korea\\
       E-mail: \email{jihnekim@gmail.com}}
 
\abstract{ 
In this talk, I discuss the mere 5\,\% of atoms in the cosmic energy pie. It is basically the chiral matter problem.  Then, I review the chiral matter problem from a grand unification (GUT) point of view, and point out that anti-SU($N$), easily implementable in string compactification, is a possibility. Here `anti-' means that the GUT group breaking is through the anti-symmetric tensor field(s), e.g. $\langle \Phi^{[AB]}\rangle= \langle\Phi_{[AB]}\rangle$.  We argued for an anti-SU(7) $( = SU(7) \times U(1))$ families-unification model from a $\Z_{12-I}$ orbifold compactification. The $\Z_{12-I}$ orbifold is briefly discussed and the multiplicity 2 in the $T_3$ twisted sector is the key obtaining three chiral families. 
Yukawa coupling structure is shown to be promising. A numerical study shows that the  anti-SU($7$) model satisfies the CKM fit. It is shown that the doublet-triplet splitting is obtained naturally, where the dominant process for proton decay is by the exchange of GUT scale gauge bosons such that $p\to \pi^0+e^+$ is the dominant channel.  
}

\FullConference{18th International Conference From the Planck Scale to the Electroweak Scale \\
                 25-29 May 2015\\
                 Ioannina, Greece}

\begin{document}

\section{Introduction}

\begin{figure}[!t]
\centerline{\includegraphics[width=0.7\textwidth]{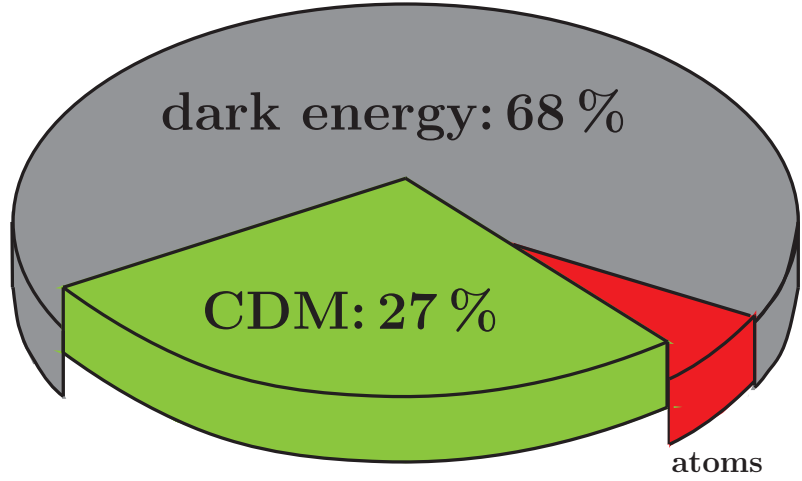} }
\caption{The cosmic energy pie.}\label{fig:EnergyPie}
\end{figure}

Recently, I paid much attention on the nature of cold dark matter(CDM) and dark energy(DE) \cite{Baer15prp,BCMRev} which constitute  27\% and 68\%, respectively, in the current cosmic energy pie \cite{PlanckCoParameters}, viz. Fig. \ref{fig:EnergyPie}.  The rest is composed of atoms, neutrinos, etc., which constitutes only 5\%.  CDM as bosonic coherent motions \cite{Preskill83} is of particular interest because the `strong CP problem' might be solved within this scheme \cite{KSVZ,DFSZ}, employing the Peccei-Quinn symmetry \cite{PQ77}. 

In this talk,  however, I discuss the mere 5\% of the cosmic energy pie from a grand unification(GUT) point of view. In particle physics language, it is the problem of surviving chiral matter at a high energy scale, usually at a GUT scale. A solution of the `chiral matter problem' is the first step achieving our presence on Earth via Sakharov's conditions, including the baryon number violation and the CP violation \cite{Kolb83}. The most familiar example is the chiral fields in the Georgi-Glasow(GG) SU(5) GUT \cite{GG74}, which is usually extended to an SO(10) GUT with the spinor representation {\bf 16} \cite{SO10}. My work related to SO($N$) chiral matter was presented a long time ago \cite{Kim80prl}. The chiral matter solution must produce a correct texture for the CKM matrices \cite{Cabibbo63,KM73,PMNS} to satisfy the low energy phenomenology \cite{CKMPDG15,PDG15}.

In the last three decades, however, unification of GUT families have not gotten much attention, because the advent of standard-like models did not need a GUT \cite{IKNQ}. Recently, we found an extended GUT, unifying three GUT families \cite{KimJHEP15}. It seems that this is the only plausible extended GUT among all low energy models from string theory (suggested so far).

The family unification might be achieved in an SO($4n+2$) chain with $n>2$, e.g. SO(18) for example. But, if one requires standard charges for quarks and leptons, they lead to vectorlike representations below a GUT scale and a solution of the chirality problem is not achieved.  On the other hand,  if one allows non-standard charges for quarks and leptons, it is possible to have a chiral solution as done in \cite{Kim80prl}. So, classification of quarks and leptons according to SO(10), but not larger than SO(10), seems promising; then the symmetry might be SO(10)$\times G_f$ where $G_f$ is a flavor group. Introduction of $G_f$ is possible in 4 dimensional(4D) field theory from a bottom-up approach. With extra dimensions, it is also possible but in terms of the standard model(SM) setup \cite{ABC03}. Being a SM at low energy, there are many couplings to be considered. To reduce the number of couplings, a symmetry must be introduced. 

Not to introduce so many couplings as in SM$\times G_f$, an extended GUT is one plausible extension. When the extended GUT is broken down to a GUT, string compactification prefers the flipped-SU(5) \cite{Barr82,DKN84}. This is because of the difficulty in obtaining an adjoint representation of SU(5) which is needed to break the SU(5) GUT down to the SM.

\section{Flipped-SU(5) or anti-SU(5), and anti-SU(N)}
\label{sec:flippedsu5}

The  minimal rank unification of 15 SM chiral fields, $\{u^c_L,u_L,d_L,e^c_L\} \oplus\{d^c_L,\nu_{e\,L},e_L\}$, is due to Georgi and Glashow(GG) under the rank 4 GUT, \GGsu5 \cite{GG74}. Any SM singlet fields must carry $\Qem=0$ since the electromagnetic charge is fully embedded in SU(5). The 15 SM fields can form a spinor representation {\bf 16} of SO(10) with one more two-component neutral fermion $N_L^c$, making up , $\{u^c_L,u_L,d_L,e^c_L\} \oplus\{d^c_L,\nu_{e\,L},e_L\} \oplus N_L^c$. This SO(10) is a rank 5 group. Branching of  {\bf 16} to the SM can be achieved via two routes, one by the GG route,  \GGsu5$\times$U(1), and the other by the flipped-SU(5) route, \flip$\equiv$SU(5)$'\times$U(1)$_X$. Under \flip, the 16 chiral fields are grouped into  $\{d^c_L,u_L,d_L,N^c_L\} \oplus\{u^c_L,\nu_{e\,L},e_L\}\oplus e^c_L$. Since $e^c_L$ and $N^c_L$ are exchanged (and also $u^c_L\leftrightarrow d^c_L$), from the GG grouping, the \flip\,route is called `flipped-SU(5)' \cite{Barr82}. It can be called `anti-SU(5)' \cite{DKN84} because of the fact that the GUT breaking can be achieved by a vacuum expectation value(VEV) of an anti-symmetric tensor-field {\bf 10} which contains an electromagnetically neutral field, \ie $N_L^c$. But, \GGsu5 and \flip~are not models of family unification. To unify families, we must introduce an extra family symmetry such that a full group may be \GGsu5$\times G_f$ or \flip$\times G_f$. Instead of the factor group $G_f$, families can be unified in  extended GUTs of SU($N$), including \GGsu5  or \flip~as its subgroups \cite{Georgi79, Frampton79, Kim80prl}. If SU($N$) is further extended to rank $N=2n+1$ group SO($4n+2$), the GG route does not produce any chiral family. On the other hand, the \flip~route of SO($4n+2$) can produce chiral families. 

The tensor-indices notation of SU($N$) representation is useful in calculating the U(1) charges. These are $\Psi^{[A]},\Psi^{[AB]},\Psi^{[ABC]},$ etc., where $A=1,2,\cdots N$, and indices inside [~~] are antisymmetrized. For example, $\one+\Psi_{[A]}+\Psi^{[AB]}+\Psi_{[ABC]}$ for $N=7$ constitutes the spinor representation {\bf 64} of SO(14) \cite{Kim81}. {\bf 64} does not produce any family of the SM if the GG route is chosen. In the GG language, it is vector-like, containing two L-handed doublets and two R-handed doublets both for quarks and leptons.
 On the other hand, if the anti-SU(7) route is taken, it produces chiral fields of three lepton families and two quark families \cite{Kim80prl}. This can be easily seen that for the two R-handed doublets, the $\Qem$ charges are one unit raised and lowered, respectively. Thus, we have the following doublets in the anti-SU(7) route of SO(14), 
 \begin{eqnarray}
\left( \begin{array}{c} \nu_e\\ e\end{array}\right)_L ,  \left( \begin{array}{c} \nu_\mu\\ \mu\end{array}\right)_L ,  \left( \begin{array}{c} \tau^+\\ \nu_\tau^c\end{array}\right)_R ,  \left( \begin{array}{c} M^-\\ M^{--}\end{array}\right)_R ,  \left( \begin{array}{c} u\\ d\end{array}\right)_L ,  \left( \begin{array}{c} c\\ s\end{array}\right)_L ,  \left( \begin{array}{c} q_2^{5/3}\\ t\end{array}\right)_R ,  \left( \begin{array}{c} b\\ q_1^{-4/3}\end{array}\right)_R ,  \label{eq:flip7}
\end{eqnarray}
which contains three SM L-handed lepton doublets ($(\tau^+,\nu_\tau^c)_R^T$ can be written in terms of their antiparticles as $(\nu_\tau,\tau)_L^T$) and two SM L-handed  quark doublets. Among 64, the remaining 32 fields are SU(2)$_W$ singlets. $t$ and $b$ in the doublets are not in the same doublet, and hence this model is ruled out. In addition, there are non-standard charges such as the doubly charged lepton and $\Qem=\frac53,-\frac43$ quarks, and hence the weak mixing angle at the GUT scale is $\frac{3}{20}$.\footnote{The weak mixing angle problem, however,  can be fitted using another scale present in the model \cite{Kim80prl}.} Nevertheless, this anti-SU($N$) route shows the survival of chiral families from SO($4n+2$) for $n\ge 3$.

Our objective here is to have three SM families,
\begin{eqnarray}
\left( \begin{array}{c} \nu_e\\ e\end{array}\right)_L ,  \left( \begin{array}{c} \nu_\mu\\ \mu\end{array}\right)_L ,  \left( \begin{array}{c} \nu_\tau\\ \tau\end{array}\right)_L ,    \left( \begin{array}{c} u\\ d\end{array}\right)_L ,  \left( \begin{array}{c} c\\ s\end{array}\right)_L ,  \left( \begin{array}{c} t\\ b\end{array}\right)_L ,~~  {\rm SM~charged~singlets}. \label{eq:ThreeSMf}
\end{eqnarray}
If there is a desert from the electroweak scale up to a GUT scale, evolution of couplings from TeV scale to the GUT scale unifies, leading to  $\sin^2\theta_W\simeq\frac38$ \cite{GQW74}. After the heterotic string compactification, this field theoretic extension of GUT models has not gotten much attention.  This is because string compactification can contain the GUT breaking mechanism intrinsically, leading to standard-like models \cite{IKNQ}. But, the weak mixing angle problem is serious \cite{mixAngleSt} and it is better to have a GUT at the unification point as
\begin{equation}
\left. \sin^2\theta_W\right|_{E\simeq 2.5\times 10^{16\,}\gev}~\simeq~\frac38,~{\rm by~ evolution~of~gauge~couplings~up~from~}M_W.
\label{MixAngleatGUT}
\end{equation}
If some non-standard charges appear above this unification scale, $\sin^2\theta_W$ above that scale need not be $\frac38$, which might be the case due to the extra U(1) factor in anti-SU($N$). GUTs containing SU(5)$\times$U(1), anti-SU(5), are automatic solutions to the  weak mixing  angle problem due to the standard charges below the GUT scale. Thus, the question is, ``Can we succeed in finding (\ref{MixAngleatGUT}) in a unification of GUT families from string?'' Below, I present a possibility.

The model should not be an anti-SU($N$) subgroup of SO($4n+2$) as commented above. So, the embedding chain of the flipped SU(5) is
\begin{equation}
\rm SU(5)\times U(1)_X \in anti\,SU(7)\in E_8,
\end{equation}
but not in the chain of
\begin{equation}
\rm SU(5)\times U(1)_X \in SO(10)\in SO(14) \in E_8.
\end{equation}
If the flipped-SU(5) is the GUT group, it is easy to obtain them in compactifications of superstring models \cite{Ellis89,KimKyae07}.

Unification of GUT families in an extended GUT SU($N$), one must use anomaly free representations. Anti-SU($N$) has an additional factor U(1)$_Z$, but a prerequiste is not to have the nonabelian anomaly. The number of families is counted by the number of $\Psi^{[ab]}$ in the \flip~subgroup where $a$ and $b$ are the SU(5) indices. The SU($N$) anomalies of the representation $[n]$ with $n$ antisymmetric indices are
\begin{eqnarray}
 &&~{\cal A}([m])= \frac{(N-3)!(N-2m)}{(N-m-1)!(m-1)!},\\ 
 &&~{\cal A}({[1]})=1,~{\cal A}\left({ [2]}  \right)=N-4,~ {\cal A}\left([3] \right) = \frac{(N-3)(N-6)}{2},\,{\rm etc.} \label{eq:anomaly}
\end{eqnarray}
where ${\cal A}([{N-m}])=-{\cal A}([m])$.
Thus, the number of families, counted by $[n]\ge [2]$, are
\begin{eqnarray}
{\rm SU(5)}&:& [2]\to n_f=1\\
 {\rm SU(6)}&:& [3]\to n_f=0, ~[2]\to n_f=1\\
 {\rm SU(7)}&:&[3]\to n_f=1, ~[2]\to n_f=1\\
 {\rm SU(8)}&:& [4]\to n_f=0, ~[3]\to n_f=2, ~[2]\to n_f=1\\
 {\rm SU(9)}&:& [4]\to n_f=5, ~[3]\to n_f=3, ~[2]\to n_f=1\\
 {\rm SU(11)}&:& [5]\to n_f=-5, ~ [4]\to n_f=9, ~[3]\to n_f=5, ~[2]\to n_f=1. \label{eq:NumberGGf}
\end{eqnarray}
The  representations [1] and [2], for example, are identified as antisymmetric tensor fields of
\begin{eqnarray}
[1]\equiv \Phi^{[A]}=\left(
\begin{array}{c}
\alpha_1\\[0.2em] \alpha_2\\[0.2em]  \alpha_3\\[0.2em]  \alpha_4\\[0.2em]  \alpha_5\\[0.2em]  f_6\\ \cdot \\[0.2em]  \cdot\\[0.2em]    \cdot\\[0.3em]  f_N\\[0.2em]
\end{array}\right),\hskip 0.5cm
[2]\equiv \Phi^{[AB]}=\left(\begin{array}{cccccccc}
0, &\alpha_{12}, &\cdots,&\alpha_{15} & {\Big|}&\epsilon_{16},&\cdots,&\epsilon_{1N}\\[0.2em]
-\alpha_{12},& 0,&\cdots,&\alpha_{25}&{\Big|}&\epsilon_{26},&\cdots,&\epsilon_{2N}\\ 
\cdot&  \cdot&  \cdot&  \cdot&{\Big|}&  \cdot& \cdot& \cdot \\ 
\cdot&  \cdot&  \cdot&  {\color{red}\alpha_{45}}&{\Big|}&  \cdot& \cdot& \cdot \\ 
-\alpha_{15}, &-\alpha_{25},&  \cdots,&0&{\Big|}&\epsilon_{56},&\cdots,&\epsilon_{5N}\\[0.2em]
\hline
-\epsilon_{16},& -\epsilon_{26}\,&\cdots,&-\epsilon_{56}&{\Big|}&0,&\cdots,&\beta_{6N}\\ 
\cdot&  \cdot&  \cdot&  \cdot&{\Big|}&  \cdot& \cdot& \cdot \\ 
-\epsilon_{1N},& -\epsilon_{2N}\,&\cdots,&-\epsilon_{5N} &{\Big|}&-\beta_{6N},&\cdots,&0
 \end{array}
 \right) \label{eq:FundSUN}
\end{eqnarray}
where the red colored $\langle\Phi^{[45]}\rangle$ separates color($\alpha=1,2,3$) and weak($i=4,5$) indices.
Note that the smallest number of non-singlet fields is possible in anti-SU(7),
\begin{equation}
\Psi^{[ABC]}\oplus 2\,\Psi^{[AB]}\oplus 8\,\Psi_{[A]}\oplus{\rm Singlets}.
\end{equation}
We will obtain this minimal model from the compactification of $\EE8$ heterotic string on a $\Z_{12-I}$ orbifold \cite{KimJHEP15}. Obtaining the spectra from string is very important because the resulting spectra are free of gravitational obstruction of discrete and global symmetries \cite{KraussWilczek,Ibanez92,Barr92}. 
  
\section{Discrete symmetries: Mother of global symmetries}
\begin{figure}[!t]
\centerline{\includegraphics[width=0.5\textwidth]{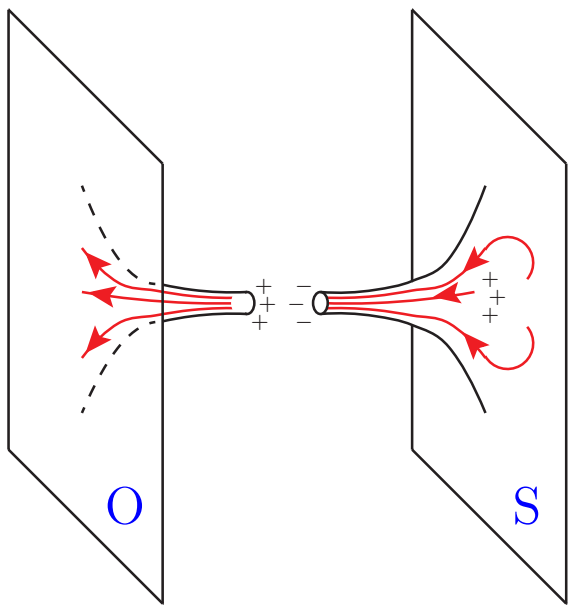}
 }
\caption{The visible Universe O is connected to a shadow world by a wormhole \cite{Coleman88}.}\label{Fig:Wormhole}
\end{figure}

It is known for some time that gravity does not respects global symmetries. On the other hand, topology changes of the metric cannot break gauge symmetries \cite{Giddings88,Coleman88}. We show a cartoon for this in Fig. \ref{Fig:Wormhole}, where our Universe (denoted as O) is connected to a shadow world (denoted as S) by the topology change through a wormhole.
Some gauge field lines are also shown as red flux lines in  Fig. \ref{Fig:Wormhole}.
If some + U(1) gauge charges are flown through the wormhole to S, the flux lines are dragged behind in O through the wormhole. The observer O will see interactions in the visble Universe by cutting off the wormhole, \ie by cutting the flux lines in the wormhole. Because of the gauge symmetry, the cutted surfaces are charged as + and -- as shown in 
 Fig. \ref{Fig:Wormhole}. Namely, the net charge flown to S is zero. Or, in other words the Observer O can recognize  how much charges are lost by looking at the fluxes going into  the wormhole without cutting off the wormhole. Due to the long range flux lines, gravity cannot spoil gauge symmetries. 
 
On the other hand, spontaneously broken global symmetries do not have the corresponding long range flux lines and the above argument is invalid. So, gravity spoils the ideas of  spontaneously broken global symmetries. In particular, axions from the PQ symmetry is in danger since gravity does not allow the global symmetry below the Planck scale \cite{Barr92}. If some fine-tuning is allowed for the spontaneously broken PQ global symmetry, the terms in the potential with dimension less than 8 is required to be absent. It is a severe tuning because $(10^{11\,}\gev/10^{18\,}\gev)^7\approx 10^{-49}$ \cite{Barr92}. In the bottom-up approach, even some discrete symmetries are in danger. If a discrete symmetry such as CP \cite{Nelson84} or SUSY R-parity \cite{Hall83} is required in a theory, it is better for them to arise as a subgroup of spontaneously broken gauge symmetry, which has been named as {\it discrete gauge symmetry} \cite{KraussWilczek,Ibanez92}.

\begin{figure}[!b]
\centerline{\includegraphics[width=0.4\textwidth]{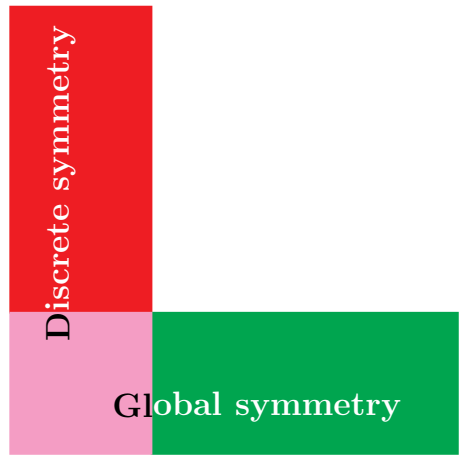}
 }
\caption{A cartoon for the discrete and global  symmetries. The terms in the red and lavender parts respects discrete symmetries arising from string compactification, and the terms in  the green and lavender parts respect a global symmetry.}\label{Fig:discrete}
\end{figure}

In the top-down approach of string compactification, one needs not consider this gravity problem on discrete symmetries because string theory is claimed to give a consistent gravity theory. In this regard, we note that the R-parity was constructed first in the $\Z_{12-I}$ orbifold compactification \cite{Kyae07}, and also in the $\Z_{6-II}$ orbifold compactification \cite{Ratz08}.
   
The top-down approach in string compactification introduces global symmetries under the name of {\it approximate global symmetries} \cite{IWKim07} or {\it accidental grobal symmetries} \cite{ChoiKS09}.  The situation is shown in Fig. \ref{Fig:discrete}, where the effective terms according to string-allowed discrete symmetries can have some global symmetries if we keep only a few terms, shown as the lavender part among the vertical (string allowed) interaction terms.
The terms in the  vertical column represent exact symmmetries such as anomaly-free gauge symmetries and string allowed discrete symmetries. If we consider only a few terms in the lavender part, we can talk about a {\em global symmetry}. For such a global symmetry, we depicted it as the  green part in Fig. \ref{Fig:discrete}. The global symmetry is broken by the terms in the red part.
\begin{figure}[!t]
\centerline{ \includegraphics[width=0.35\textwidth]{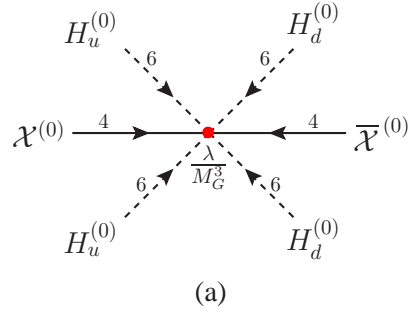} 
}
\centerline{(a)}\vskip 1cm
\centerline{ \includegraphics[width=0.52\textwidth]{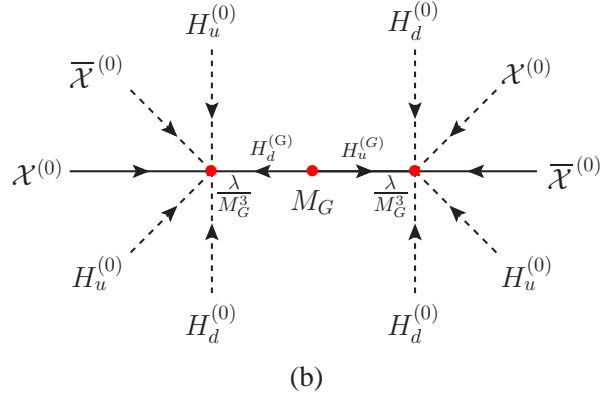}
}
\centerline{(b)}\vskip 0.2cm
\caption{Feynman diagrams for \Ude~symmetry: (a) the defining diagram of \Ude~corresponding to the lavender part of Fig. 3, and (b) the \Ude~breaking diagram corresponding to the red part of Fig.3.  }\label{Fig:FeynmanGlobal}
\end{figure}
The most studied global symmetry is the Peccei-Quinn (PQ) symmetry \UPQ\,\cite{PQ77}.  
Note that the global symmetry violating terms belong to the red part in Fig. \ref{Fig:discrete}. For the QCD axion, the dominant breaking term is the QCD anomaly term, which leads to the QCD axion mass in the range of milli- to nano-eV for $f_a\simeq 10^{9-15}\,\gev$. In the BEH portal scenario, DE  may be interpreted as the vacuum energy of a pseudoscalar $\varphi$ with its coherent motion not commenced yet. For this temporary cosmological constant,  the psudoscalar mass is in the range $10^{-33}\sim 10^{-32}\,\eV$ \cite{Carroll98}. The QCD anomaly term is too large to account for the DE scale of $10^{-46\,}\gev^4$, and we must find out a QCD-anomaly free global symmetry. It is possible by introducing two global U(1) symmetries  \cite{KimNilles14,KimJKPS14}.
 In the  temporary cosmological constant scenario, the QCD anomaly should be absent, which implies that the QCD axion must be considered together as in \cite{KimPRL13,KimNilles14,KimJKPS14}. 
An example based on the flipped-SU(5)$\times$SU(5)$'$ model \cite{KKHuh09} from $\Z_{12-I}$ orbifold compactification uses $\Z_{10\,R}$ \cite{KimPLB13}. In Fig. \ref{Fig:FeynmanGlobal}\,(a), corresponding to the lavender part of Fig. \ref{Fig:discrete}, we present the \Ude~defining diagram, which is already suppressed by $\Mp$.  In Fig. \ref{Fig:FeynmanGlobal}\,(b), corresponding to the red part of Fig. \ref{Fig:discrete}, we present the \Ude~breaking diagram, which is suppressed by $\Mp^3$.  Because the discrete symmetry require the presence of $(H_uH_d)^2$, the DE pseudo-Goldstone boson mass is extremely small \cite{KimNilles14,KimJKPS14}. 
      
       On the other hand, if $\varphi$ has oscillated already, it might have worked as an ultra light axion (ULA).  For $m_{\varphi}\simeq 10^{-22\,}$eV and $f_\varphi\simeq  M_{\rm GUT}$, the cusp/core and the dwarf galaxy problems can be understood  \cite{MarshSilk14,Chiueh14}. With an approximate global symmetry from an exact discrete discrete symmetry, this scenario can be also realized \cite{MarshKim15}. 
       
       In these temporary cosmological constant and the ULA scenarios, the QCD anomaly should be absent, which implies that the QCD axion must be considered together as in \cite{KimPRL13,KimNilles14,KimJKPS14,MarshKim15}. Thus, global symmetries from discrete symmetries of string compactification are useful in particle physics, involving QCD, and cosmology because they would be free from the gravity spoil \cite{Barr92}. We do not consider the SU(2)$_W$ anomaly in these considerations because the non-perturbative effect due to SU(2)$_W$, $e^{8\pi^2/g^2_{\rm weak}}$, is completely negligible.
 
\section{Unification of GUT families from $\Z_{12-I}$}
 
 \subsection{Requirements for families-unified GUT}
Therefore, we pursue the unification of GUT families from string compactification, for which  there has not been any model before  Ref. \cite{KimJHEP15}. Being a GUT model, we must fulfil the following requirements:
\begin{itemize}
\item Realization of the  
      doublet-triplet splitting.   
\item   The  existence of a GUT scale scalar, breaking  the GUT group down to the SM, via the BEH mechanism. 

\item An acceptable fermion mass matrix, which was the original motivation for the GUT families-unification.
\end{itemize}
The difficulty of the doublet-triplet splitting problem is easy to understand in the SU(5) breaking pattern of the SUSY SU(5). The superpotential $W$ in the SUSY SU(5) \cite{DimGeorgi81,Sakai81} is given with the adjoint BEH field {\bf 24} such that
\begin{equation}
\langle W\rangle=M_1\langle\, \fiveb_{\rm BEH} \five_{\rm BEH}\rangle+
\left\langle \fiveb \left|\left(\begin{array}{ccccc}
v_c& 0&0&0&0\\
0&v_c&0&0&0\\
0&0&v_c&0&0\\
0&0&0&-\frac32 v_c&0\\
0&0&0&0&-\frac32 v_c\end{array} \right)
\right|\five \right\rangle
\end{equation}
where a fine-tuning is needed with the {\bf 24} BEH boson,
\begin{eqnarray}
M_1+\frac32 v_c\to 0.\label{eq:finetuning}
\end{eqnarray}
In SUSY GUTs, therefore the doublet-triplet solution is very important as noticed by Kawamura's 5D SUSY SU(5) which attracted a great deal of attention \cite{Kawamura01}. In this 5D model, the BEH fields $H_u$ and $H_d$ live in the bulk. The Kaluza-Klein modes  $H_u^{n}$ and $H_d^{n}$ is proportinal to $2n/R_c$ where $R_c$ is the compactification size. Thus, we obtain a massless pair of BEH doublet superfields for $n=0$. This simple 5D orbifold  example is a special case of string orbifolds, known from 
1987 \cite{IKNQ}.   

\begin{figure}[!t]
\centerline{ \includegraphics[width=0.75\textwidth]{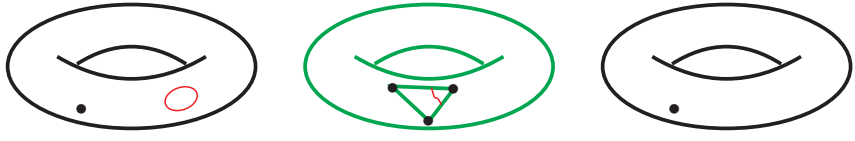} 
}
\caption{Three two-tori. In $\Z_{12-I}$ orbifolds, Wilson lines can distinguish three fixed points in the second torus. }\label{Fig:TriTori}
\end{figure}

The orbifolds use the discrete symmetries to compactify six internal dimensions. The six internal dimensions are viewed as the direct product of three tori. Some discrete symmetries $\Z_N$ with the twist $\phi_s$ and the number of fixed points are shown in Table \ref{tab:lattices}. 
\begin{table}[!h]
\begin{center}
\begin{tabular}{|c|c|c|c|}
 \hline &&&\\[-1.1em]
Order $N$ &   $\phi_s$  & $\phi_s^2$ &No. of fixed points\\[0.2em]
 \hline &&&\\[-1.1em]
  4 &   $\frac14(2~1~1)$ &$\frac{3}{8}$ & 16\\[0.2em] \hline &&&\\[-1.1em] 
  6-I  &$\frac16(2~1~1)$ &$\frac{1}{6}$& 3\\[0.2em] \hline &&&\\[-1.0em]
   6-II  & $\frac16(3~2~1)$&$\frac{14}{36}$ & 12\\[0.2em] \hline &&&\\[-1.1em] 
  8-I & $\frac18(3~2~1)$&$\frac{14}{64}$ & 4 \\[0.2em]\hline &&&\\[-1.1em]
 8-II  & $\frac18(4~3~1)$ &$\frac{26}{64}$& 8\\[0.2em] \hline &&&\\[-1.1em]
12-I  & $\frac1{12}(5~4~1)$&$\frac{42}{144}$ & 3\\[0.2em] \hline &&&\\[-1.1em] 
12-II  & $\frac1{12}(6~5~1)$&$\frac{62}{144}$ & 4 \\[0.2em]
\hline
\end{tabular}
 \caption{Number of fixed points of some non-prime orbifolds.}\label{tab:lattices}
\end{center} 
\end{table}
In the $\Z_{12-I}$ orbifold, there is only one fixed point in the first and third tori, but there are three fixed points in the second torus, which is depicted in Fig. \ref{Fig:TriTori}. In the second torus, three fixed points can be distinguished by Wilson lines. 

Another requirement on the BEH field for the GUT breakingis easily achieved with the anti-SU(5) \cite{Ellis89,KimKyae07}. For a unification of GUT families, therefore, an anti-SU($N$) is the first try to consider.

The last (but not the least) requirement is that the fermion mass texture must be realized in this unification of families. After all, the prime objective in the unification of families is to understand the similarity and the difference among families. The similarity is the gauge interaction. The difference is the Yukawa interaction which are the sources for the mass hierarciy among families and flavor mixings. Since families with quarks and leptons are treated universally, there are fewer Yukawa couplings than in models without families-unification and the fermion mass texture will be simple enough to be studied in detail \cite{KimCKMGUT,KimCKMPMNS}.

 \subsection{Identification of massless fields in $\Z_{12-I}$ orbifold}
 
Most successful string compactification appllicable to particle phenomenology comes from heterotic string, mainly because the large gauge groups, $\EE8$ and SO(32), are already given. The heterotic string has not given the moduli stabilization explicitly, but the scheme must be there if its dual theory possesses the solution. For the phenomenologically useful compactifications, there are so many fields and it is practically impossible to find out its dual theory spectrum. But, we anticipate that in its dual form the issue of moduli stabilization would be resolved.

The orbifold spectra are broadly composed of two parts, one in the untwisted sector $U$ and the other in the twisted sector $T$. The untwisted states are closed strings and twisted states are sitting at the fixed points, which are shown as red curves in the first and second tori in Fig. \ref{Fig:TriTori}, respectively. Now, we present the massless spectra.

The model we discuss below is based on the following shift vector and Wilson lines \cite{KimJHEP15}
\begin{eqnarray}
V^a=\left\{\begin{array}{l} V_0=\left(\frac{-5}{12},\,\frac{-5}{12},\,\frac{-5}{ 12},\,\frac{-5}{12},\,\frac{-5}{12},\,\frac{-5}{12},\,\frac{-5}{12} ;\, \frac{+5}{12}\right)
\left(\frac{4}{12},\, \frac{4}{12},\, \frac{4}{12},\,\frac{4}{12},\, 0,\, \frac{4}{12},\, \frac{7}{12},\, \frac{ 3}{12}\right)',~V^2_0=\frac{ 338}{144},\\[0.2em]
a_3=a_4=(\frac{1}{3},\,\frac{1}{3},\,\frac{1}{3},\,\frac{1}{3},\,\frac{1}{3},\,\frac{1}{3},\,\frac{1}{3};\frac{1}{3}) (\frac{-1}{3},\,\frac{-1}{3},\,\frac{-1}{3},\,\frac{-1}{3},\,0,\,0,\, \frac{5}{3},-1)' .  \end{array}\right. \label{eq:Model}
\end{eqnarray}

\subsubsection{Untwisted sector $U$}
 
 To count the number of families, the first thing to check is to count the number of $\ten$'s under the flipped-SU(5). We require $\Psi^{[ABC]}_{L,R}$ to appear in the untwisted sector $U$ because it cannot be obtained in the $T$ sectors. The possibilities of $\tenb_{L,R}$ from $U$ are shown in Table \ref{tab:tenfromU} together with the shift vector conditions and the corresponding chiralities.
 \begin{table} [!h]
\begin{center}
\begin{tabular}{|c|c|c|c| c| }
 \hline  
$U_i$ &\,Number~of\,${\bf 10}\,\rm s$\,&\,Tensor~form\,&\,Chirality\,&\,$[p_{\rm spin}]~(p_{\rm spin}\cdot\phi_s)$\, \\[0.1em] \hline  
   ~$U_1\,(p\cdot V=\frac{5}{12})$~  & 1  &~$\Psi^{[ABC]}$~ &R &$[\oplus;+++]~~\left(\frac{+5}{12} \right)$  \\   
    $U_2\,(p\cdot V=\frac{4}{12})$ & 3 &   ~$\Psi^{[ABC]}$~&L &$[\ominus;++-]~~\left(\frac{+4}{12} \right)$    \\ 
    $U_3\,(p\cdot V=\frac{1}{12})$ &  1  &~$\Psi^{[ABC]}$~ &L  &$[\ominus;+-+]~~\left(\frac{+1}{12} \right)$   \\[0.1em]  
\hline
\end{tabular}
\end{center}
\caption{The number of $\ten$'s from  anti-SU(7)~in ${\bf Z}_{12-I}$. }
\label{tab:tenfromU}
\end{table}
Thus, the model (\ref{eq:Model}) gives one $\Psi^{[ABC]}_R$, satisfying $p\cdot V=\frac{5}{12}$.
 
\subsubsection{Twisted sector $T$}

\begin{figure}[!t]
\begin{center}
\includegraphics[width=0.4\textwidth]{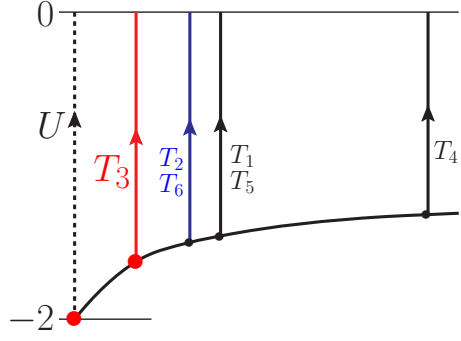} 
\end{center}
\caption{The $\Z_{12-I}$  raise of the vacuum energy in the left movers: for 
$U({\rm by\,}2)$ and $T_k({\rm by\,}2\tilde{c}_k)$.} \label{fig:Raise}
\end{figure}

The massless spectra in the twisted sectors must satisfy the spectra conditions. The widely discussed $\Z_{12-I}$  \cite{LNP696, KimKyae07} and $\Z_{6-II}$ \cite{Raby05} have the following vacuum energy contributions, 
\begin{eqnarray}
&\Z_{12-I}:& \left\{\begin{array}{ll}
2\tilde{c}_k:  &
 ~ \frac{210}{144}(k=1),~ \frac{216}{144}(k=2),~ {\color{red}\frac{234}{144}(k=3),}~ \frac{192}{144}(k=4),~ \frac{210}{144}(k=5),~ \frac{216}{144}(k=6),\\[0.3em]
 2 {c}_k: &
  ~ \frac{11}{24}(k=1),~ \frac{1}{2}(k=2),~ \frac{5}{8}(k=3), ~\frac{1}{3}(k=4),~ \frac{11}{24}(k=5),~ \frac{1}{2}(k=6), 
 \end{array}\right.\label{eq:Twist121}\\
&\Z_{6-II}:& \left\{\begin{array}{ll}
2\tilde{c}_k:  &
 ~ \frac{25}{18}(k=1),~ \frac{14}{9}(k=2),~ \frac{3}{2}(k=3),\\[0.3em]
 2 {c}_k: &
  ~ \frac{7}{ 18}(k=1),~ \frac{5}{9}(k=2),~ \frac{1}{2}(k=3),  \end{array}\right.\label{eq:Twist62}
  \end{eqnarray}
where $\tilde{c}_k$ and $c_k$ are for the L- and R-sectors, respectively. In Fig. \ref{fig:Raise}, we show schematically how the L-contributions $\tilde{c}_k$ raise the tachyonic left-mover states to the massless ones. Similarly,  R-contributions $c_k$ raise the tachyonic right-mover states to massless ones. For the  $\Z_{12-I}$ orbifold, the raise given in Eq. (\ref{eq:Twist121}) is shown pictorially in Fig. \ref{fig:Raise}. The raise 2 in the untwisted sector is just the root lattice amount in obtaining  the unbroken gauge group, $P^2=2$. Among the twisted sectors, as one can see from Eqs. (\ref{eq:Twist121}) and (\ref{eq:Twist62}), the largest raise is at the $T_3$ sector of   $\Z_{12-I}$. This is just the right amount to have {\bf 21} in anti-SU(7). There is no other possibility of the unified GUT families from the twisted sectors in Eqs. (\ref{eq:Twist121}) and (\ref{eq:Twist62}).
 
\begin{figure}[!b]
\centerline{ \includegraphics[width=0.35\textwidth]{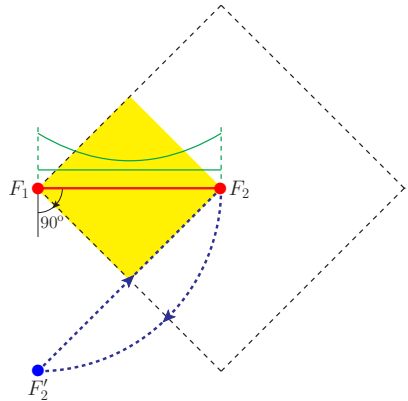} 
}
\caption{The $\Z_4$ fixed points $F_1$ and $F_2$ on a two-torus. }\label{Fig:Zfour}
\end{figure}

As one can see from Fig. \ref{Fig:TriTori}, there are three fixed points in the second torus which can be distinguished by Wilson lines. These distinguished fixed points correspond to $V, V+a_3,$ and $ V-a_3$. $\phi_s$ of  $\Z_{12-I}$
in Table \ref{tab:lattices} cannot be used for distinguishing the different points in $T_3$ and $T_6$, because $3a_3$ are composed of integers. Thus, at $T_3$ and $T_6$, we just calculate the spectrum and use the multiplicity of the discrete symmetry $\Z_4$ since $3V$ is composed of integer multiples of $\frac14$. At $T_3$ and $T_6$, we require the gauge invariance condition because the Wilson line is not useful. The same gauge invariance condition is applied in the untwisted sector also. In the second torus, the $\Z_4$ symmetry allows two fixed points as shown in Fig. \ref{Fig:Zfour}.
In the torus, a $\Z_4$ rotation by 90 degrees moves $F_2$ to  $F_2'$  (by the blue dashed-arc) which is identified as  $F_2$ by a lattice shift (by the blue dashed-line). So,  $F_2$ is a fixed point. The origin  $F_1$ is also a fixed point. In fact, the explicit calculation shows that the multiplicity is 2 \cite{KimJHEP15}. $\Psi^{[AB]}$ appears only in $T_3$ as
 \begin{eqnarray}
P=(\underline{2~2~1~1~1~1~1};~-1)(-1~-1~-1~-1~0~-1~-2~-1)', \label{eq:T3P}
  \end{eqnarray}
 \begin{eqnarray}
P+3V=\left(\underline{\frac{3}{4}~\frac{3}{4}~\frac{-1}{4}~\frac{-1}{4}~\frac{-1}{4}~\frac{-1}{4}~\frac{-1}{4}};~\frac{1}{4}\right)\left(0~0~0~0~0~0~\frac{-1}{4}~\frac{-1}{4}\right)', \label{eq:T3Gauge}
 \end{eqnarray}
where the underlined entries mean all possible permutaions and satisfies the needed condition $(P+3V)\cdot a_3=0$. Also, the masslessness condition is satisfied since $(P+3V)^2=\frac{234}{144}$ of Eq. (\ref{eq:Twist121}). The chirality is calculated as R. Since the multiplicity in $T_3$ is two, we obtain two $\Psi^{[AB]}_R$'s. Note that non-singlets have the numbers, as shown above the underline in Eq. (\ref{eq:T3Gauge}), differing by one unit. So, if there are many indices in $\Psi^{[A\cdots]}$, the masslessness condition is easily violated. For $\Psi^{[AB]}$, the condition is barely satisfied. It is barely satisfied with two $\frac34$'s and five $\frac{-1}{4}$'s. One can check that two $\frac23$'s and five $\frac{-1}{3}$'s will give a too large value. Also,  two $\frac56$'s and five $\frac{-1}{6}$'s will give a too large value. Similar comments apply to other twisted sectors.\footnote{The other interesting possibility will be $\Z_{8-II}$ where $\tilde{c}_2$ is $\frac{234}{144}$.}  This is the reason that there is no other anti-SU(7) model, allowing $\ten$'s in the  twisted sectors. 

 \begin{table}[!t]
 \scriptsize
\begin{center} 
\begin{tabular}{|c|cc|c|c|}
   \hline &&&&  \\[-1.15em]
 &~${\cal P}\times$(rep.) & Sector& ~Weight~ & ${\rm Rep\,}_{5X}$ \\[0.15em]   
  \hline
 &&&&  \\[ -1.05em]
$(a)$ & $\Psi^{[ABC]}_R$&  $U_1 $  &  $\left(\underline{----+++};+\right)(0^8)'$ & $2\,{\ten}_{1},{\tenb}_{-1}$   \\ [0.3em] \hline &&&&  \\ [-1.05em]
$(b)$ & $2\,\Psi^{[AB]}_R$ & $T_3 $  & $\left(\underline{\frac{ 3}4\, \frac{ 3}4\, \frac{ -1}4\, \frac{-1}4\, \frac{ -1}4\, \frac{ -1}4\, \frac{ -1}4}; \frac{ 1}4 \right)\left(0^6\, \frac{-1}{4}\, \frac{-1}4  \right)'$ &$2\,{\ten}_{1},4\,{\five}_{-2}$  \\[0.3em]  \hline &&&& \\[-1.05em]
$(c)$ &$~8\,\Psi_{[A] R } $ & $T_3$& $\left(\underline{\frac{ -3}4\, \frac{1}4\, \frac{ 1}4\, \frac{1}4\, \frac{ 1}4\, \frac{ 1}4\, \frac{ 1}4}; \frac{- 1}4 \right)\left(0^6\, \frac{-1}{4}\, \frac{-1}4  \right)'$ &$8\,\fiveb_{-3}, 16\,{\one}_{0}$  \\[0.3em] \hline &&&&  \\[-1.05em]
$(d)$ &$\Psi_{[A]R }$ & $T_5^+$& $\left(\underline{\frac{11}{12} \frac{-1}{12} \frac{-1}{ 12} \frac{ -1}{12} \frac{ -1}{12} \frac{-1}{12} \frac{-1}{12}}; \frac{-3}{12}\right)
\left( 0 \,0 \, 0 \, 0\,  0  \frac{4}{12} \frac{-3}{12} \frac{-3}{12}\right)'$ &${\fiveb}_{2},2\,{\one}_{-5}$   \\[0.3em] \hline &&&&  \\[-1.05em]
$(e)$ &$\Psi^{[A]}_R$ & $T_6 $& $\left(\underline{ -1, 0^6};0\right)\left(0^6,\frac{ 1}{2},\frac{ 1}{2} \right)'$ &${\fiveb}_{2},{\one}_{0}$ \\[0.1em] \hline &&&& \\[-1.05em]
$(f)$ &$40 \left(\Phi_{[A]R }+\Phi^{[A]} _{R } \right) $&  $T_3$  &$\left(\underline{\frac{- 3}4\, \frac{ 1}4\, \frac{ 1}4\, \frac{ 1}4\, \frac{ 1}4\, \frac{ 1}4\, \frac{ 1}4}; \frac{ -1}4 \right)\left(0^6\, \frac{-1}{4}\, \frac{-1}4  \right)' \oplus$~H.c. &$40\,\fiveb_{-3},  40\,{\five}_{3}, 160\,{\one}_{0}$  \\[0.3em] 
\hline &&&&\\[-1.05em]
$(g)$ &$5\left(\Phi_{[A]R}+\Phi^{[A]}_R\right)$&  $T_6$  & $\left(\underline{ -1\, 0^6}\,0\right)\left(0^6\,\frac{ 1}{2}\, \frac{1}{2} \right)'\oplus$~H.c. &$5\,{\fiveb}_{2},5\,{\five}_{-2},20\,{\one}_{0}$  \\[0.3em] 
\hline &&&&  \\[-1.05em]
$(h)$ &$ 10\left( \Phi_{[A]R }+\Phi^{[A]}_{R }\right) $&  $T_5^+$  & $\left(\underline{\frac{11}{12} \frac{-1}{12} \frac{-1}{ 12} \frac{ -1}{12} \frac{ -1}{12} \frac{-1}{12} \frac{-1}{12}}; \frac{-3}{12}\right)
\left( 0 \,0 \, 0 \, 0\,  0  \frac{4}{12} \frac{-3}{12} \frac{-3}{12}\right)'\oplus$~H.c. &$10\,{\fiveb}_{2},10\,{\five}_{-2},20\,{\one}_{-5},20\,{\one}_{5}$  \\[0.25em] 
 &&&&  \\[-1.2em] 
\hline\hline &&&& \\[-1.05em]
$(a')$ & $\Psi_{[\alpha']R} $&  $U_3$  &  $(0^8)\left(\underline{-+++};-+++\right)'$ & ${\fourb}'_{0} $   \\ [0.3em] 
\hline &&&&  \\ [-1.05em]
$(b')$ & $ \Psi^{[\alpha']}_R $&  $T_1^0$  & $\left( (\frac{1}{12})^7\,  ;\frac{-1}{12} \right)\left(\underline{ \frac{10}{12}\, \frac{-2}{12}\,\frac{-2}{12}\,\frac{-2}{12}};\frac{-6}{12}\,\frac{-2}{12}\,\frac{1}{12};\frac{-3}{12} \right)' $  &${\four}'_{-5/3}$  \\[0.3em] 
\hline &&&&  \\[-1.05em]
$(c')$ & $ \Psi_{[\alpha'] R }$&  $T_4^0$  & $\left( (\frac{-1}{6})^7\,  ;\frac{1}{6} \right)\left(\underline{ \frac{-1}{3}\, \frac{1}{3}\,\frac{1}{3}\,\frac{1}{3}};0\,\frac{ 1}{3}\,\frac{ 1}{3}\,0 \right)' $  &${\fourb}'_{10/3}$   \\[0.3em] 
\hline &&&&  \\[-1.05em]
$(d')$ & $ \Psi^{[\alpha']}_R $&  $T_5^0$  & $\left(  (\frac{1}{12})^7\,;\frac{-1}{12} \right)\left(\underline{ \frac{10}{12}\, \frac{-2}{12}\,\frac{-2}{12}\,\frac{-2}{12}};\frac{-6}{12}\,\frac{-2}{12}\,\frac{-5}{12}\,\frac{ 3}{12} \right)' $  &${\four}'_{-5/3} $   \\[0.3em] 
\hline &&&&  \\[-1.05em]
$(e')$ &$ 10\left(\Phi^{[\alpha']}_R+\Phi_{[\alpha']R}\right)$&  $T_1^0$  &  $ \left(  (\frac{1}{12})^7\,;\frac{-1}{12} \right)\left(\underline{ \frac{10}{12}\, \frac{-2}{12}\,\frac{-2}{12}\,\frac{-2}{12}};\frac{-6}{12}\,\frac{-2}{12}\,\frac{1}{12};\frac{-3}{12} \right)'\oplus $H.c. &$10({\four}'_{-5/3}+{\fourb}'_{5/3}) $   \\[0.3em] 
\hline &&&& \\[-1.05em]
$(f')$ &$ 5\left(\Phi_{[\alpha']R}+\Phi^{[\alpha']}_R\right)$&  $T_4^0$  & $\left(( \frac{-1}{6})^7\,  ;\frac{1}{6} \right)\left(\underline{ \frac{-1}{3}\, \frac{1}{3}\,\frac{1}{3}\,\frac{1}{3}};0\,\frac{ 1}{3}\,\frac{ 1}{3}\,0 \right)'\oplus $~H.c.  &$5({\fourb}'_{10/3}+{\four}'_{-10/3})$  \\[0.3em] 
\hline &&&&  \\[-1.05em]
$(g')$ &$10\left(\Phi^{[\alpha']}_R+\Phi_{[\alpha']R}\right) $&  $T_5^0$  &  $ \left( (\frac{1}{12})^7\,  ;\frac{-1}{12} \right)\left(\underline{ \frac{10}{12}\, \frac{-2}{12}\,\frac{-2}{12}\,\frac{-2}{12}};\frac{-6}{12}\,\frac{-2}{12}\,\frac{-5}{12}\,\frac{ 3}{12} \right)'\oplus $H.c.  &$10({\four}'_{-5/3}+{\fourb}'_{5/3}) $    \\[0.3em] 
\hline &&&&  \\[-1.05em]
$(h')$ &$ 7\left(\Phi^{[\alpha']}_R+\Phi_{[\alpha']R}\right)$&  $T_6 $  & $\left( 0^8\right)\left(\underline{ 1\, 0\,0\,0};0\,0\,\frac{ -1}{2}\,\frac{ -1}{2} \right)'\oplus $~H.c.  &$7({\four}_{-7} +{\fourb}_{7})$  \\[0.25em] 
\hline
\end{tabular}
\end{center}
\caption{Non-singlet spectra of the anti-SU(7) model of Ref. \cite{KimJHEP15} represented as R-handed chiral fields. H.c. means the opposite numbers of those in the same site.  } \label{tab:SU7Spectra}
\end{table}

In Table \ref{tab:SU7Spectra}, we list the non-singlet spectra of anti-SU(7) \cite{KimJHEP15}. There are the needed spectra
\begin{equation}
\Psi^{[ABC]}_R~\oplus~ 2\,\Psi^{[AB]}_R~\oplus~8\,\Psi_{[A],R}~\oplus~{\rm Singlets},
\end{equation}
which will lead to three GUT families as discussed in Sect. \ref{sec:flippedsu5}. In this anti-SU(7) model, the baryon number violating superpotential is
\begin{equation}
W\sim {\bf 21}\cdot{\bf 21}\cdot{\bf 21}\cdot{\bf 21}\cdot\overline{\bf 7}
\end{equation}
which occurs at the dimension 5 level.\footnote{Note that the SUSY SU(5) model gives the baryon number violation operator at dimension 4 superpotential $W=\ten\cdot \ten\cdot\ten\cdot\fiveb$.} So, it is negligible compared to the baryon number violating (such as $X,Y$ heavy GUT) gauge boson exchange. Thus, the leading proton decay signal is
\begin{equation}
p\to  \pi^0+ e^+.
\end{equation}
 
\section{Yukawa couplings}\label{sec:Yukawa}

\subsection{Missing partner mechanism}\label{subsec:Missing}

For the missing partner mechanism, the first one to satisfy is to forbid the  term, $\mu\,\overline{\bf 7}_{\rm BEH}\,{\bf 7}_{\rm BEH}$, the generalization of the   $\mu$ term in the MSSM, $\mu H_uH_d$ \cite{KimNilles84}. For this purpose, we must go beyond the effective field theory approach. In our model, $\overline{\bf 7}_{\rm BEH}$  housing $H_u$ is present in $T_6$, and ${\bf 7}_{\rm BEH}$  housing $H_d$ is located in $T_3$ \cite{KimJHEP15}. Therefore, the twisted sector selection rules forbid the appearance of $\Mg\,\overline{\bf 7}_{\rm BEH}\,{\bf 7}_{\rm BEH}$. If the GUT breaking VEVs of non-singlet fields sufficiently dominate SU(7) singlet VEVs,   the generalized $\mu$ term can be under control. Then, the twisted sector fields can have the following superpotential terms,
\begin{eqnarray}
&&\frac{1}{M_s}\epsilon^{ABCDEFG}
\Phi_{[AB]}\Phi_{[CD]}\Phi_{[EF]}\Phi_{[G]} \label{eq:Wmiss1}\\[0.2em]
&&\frac{1}{M_s^2}\epsilon^{ABCDEFG}\Phi_{[AB]}\Phi_{[CD]}
\Phi_{[E]}\Phi_{[F]}\Phi_{[G]}.\label{eq:Wmiss2}
\end{eqnarray}
For the GUT breaking and color/weak separation, we assign a GUT scale VEV for
\begin{equation}
\langle 0| \Phi_{[45]}|0\rangle=V~(\rm the~ red ~entry~ in~ Eq.\, (\ref{eq:FundSUN})). \label{eq:VEVgut}
\end{equation}
Inserting (\ref{eq:VEVgut}) into (\ref{eq:Wmiss1}), we obtain a term like
\begin{equation}
\epsilon^{45\alpha \beta \gamma ab} \frac{\langle\Phi_{[45]}\rangle\langle \Phi_{[ ab]}\rangle}{M_s} 
\Phi_{[\alpha\beta]}\Phi_{[\gamma]}
\end{equation}
where $\alpha,\beta,\gamma$ are color indices and $a,b$ are 6, 7. Thus, the colored particle $\Phi_{[\gamma]}$ in $\Phi_{[A]}$  obtains a SUSY mass by coupling to $\Phi_{[\alpha\beta]}$ in $\Phi_{[AB]}$. Its magnitude, $\langle\Phi_{[45]}\rangle\langle \Phi_{[ ab]}\rangle/M_s $, can be large in principle. But, the weak doublet components, $\Phi_{[4]}$ or  $\Phi_{[5]}$, cannot find a partner in $\Phi_{[AB]}$ because 4, 5 are already used in $\langle\Phi_{[45]}\rangle$. Thus, the anti-SU(7) model realized the missing partner mechanism anticipated in Ref. \cite{Dimopoulos81}.

\subsection{Top quark mass and the CKM matrix}\label{subsec:CKM}

The $u$ field is  located in $U$, (a) of Table \ref{tab:SU7Spectra}, and $u^c$ field is located at $T_3$. In the anti-SU(7) model, the $U$ sector fields have small Yukawa couplings. On the other hand, both  $\{c,t\} $ and  $\{c^c, t^c\}$ fields are located  $T_3$. Therefore, order 1 Yukawa coupling of the form ${\bf 21}(T_3)\times \overline{\bf 7}(T_3)\times\overline{\bf 7}(T_6)_{\rm BEH}$ is possible if $H_u$  in $T_6$ is not removed at the GUT scale, which is possible for a range of singlet VEV. This requires a hierarchy of scales \cite{KimJHEP15},
\begin{eqnarray}
M_s\ll  \langle{\one} (T_3)\rangle.
\end{eqnarray}
At the anti-SU(7) level, the BEH field $\overline{\bf 7}_{\rm BEH}$ is located at $T_6$.  Thus, the $t$-quark mass arises from a cubic Yukawa coupling,
\begin{eqnarray}
T^{\bf 21}_{3}T^{\overline{\bf 7}}_{3}T^{\overline{\bf 7}}_{6,\rm BEH}~( t{ \rm~mass}).\label{eq:BEH3rdt}
\end{eqnarray}
Thus, $m_t=O(v_{\rm ew})$. 

On the other hand, $H_d$ in ${\bf 7}_{\rm BEH}$ is located at $T_3$. Also, $d,s,b$ and $d^c,s^c,b^c$ are located at $T_3$.  Thus, the $b$ quark mass arises from
\begin{eqnarray}
\sim\frac{1}{M_s}\,T^{\bf 21}_{3}T^{\bf 21}_{3}T^{\bf 21}_{3,\rm BEH}T^{\bf 7}_{3,\rm BEH},\label{eq:BEH3rdb}
\end{eqnarray}
which gives a ratio to $t$-quark mass as $ O( \langle T^{ \bf 21 }_{3,\rm BEH}\rangle \langle T^{ \bf 7 }_{3,\rm BEH} \rangle/ M_s \langle T^{ \bf 7 }_{6,\rm BEH}\rangle    )$, where $ \langle T^{ \bf 21 }_{3,\rm BEH}\rangle$ is the SU(5) splitting VEV $\langle\Phi^{[67]}\rangle$. Namely, we expect $m_b/m_t\sim\frac{ \langle\Phi^{[67]}\rangle}{M_s\tan\beta}$.  

\begin{figure}[!t]
\begin{center}
\includegraphics[width=0.7\textwidth]{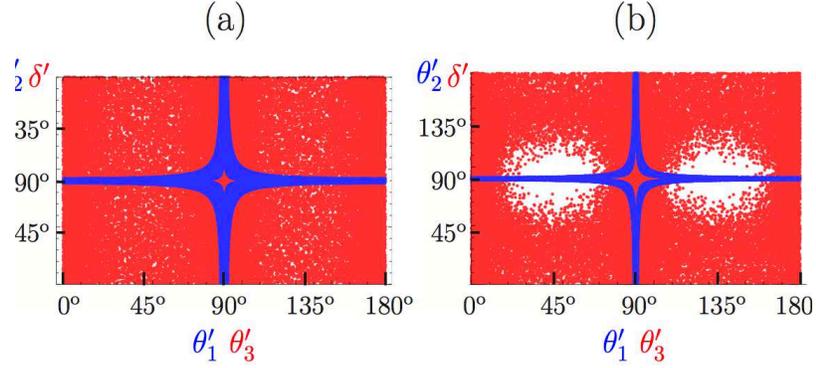} 
\end{center}
\caption{The bounds on the angles of the right-handed unitary matrix $V_R^{(u)}$ diagonalizing $M^{(u)}$, (a) $a=\frac{1}{1.5}\sin\theta_c,~b=1.5\sin\theta_c$, and (b) $a=\frac{1}{1.2}\sin\theta_C,~b=1.2\sin\theta_c$.   The white regions are not allowed.} \label{fig:Rangles}
\end{figure}

Now, we can calculate the CKM matrix from the anticipated Yukawa couplings. It does not need obtaining all SU(7) singlets in the anti-SU(7) model. One key point is that both $c $ and $t $ appear in $T_3$ but $u$ appears in $U$. Thus, the sub-matrix for $c,t$ masses has a democratic form due to the multiplicity 2 in Fig. \ref{Fig:Zfour}.
In this way, we can show that the mass matrix suggested in the anti-SU(7) model  allows the phenomenologically determined CKM matrix elements \cite{KimCKMGUT} as shown in the scatter plot, Fig. \ref{fig:Rangles}. 
$\theta_1, \theta_2,\theta_3$ are three real angles, and $\delta$ is the CP violating phase in the CKM matrix, which are determined by the $W$ and quark couplings. In Fig. \ref{fig:Rangles}, these phenomenologically determined error limits of the CKM matrix elements are used.  The coordinates 
$\theta_i'$ and $\delta'$ in Fig. \ref{fig:Rangles} are the so-called R-handed(L-handed in our chiralities of Table \ref{tab:SU7Spectra}) unitariy matrix in the process of diagonalizing the quark mass matrix.  The red colored region  does not bound the angles very much, \ie $\theta_1'$ and $\delta'$ are not constrained strongly. But the blue colored region, \ie $\theta_{1,2}'$ are  constrained. But these R-handed matrix elements cannot be probed at low energy.
 
\section{Conclusion} \label{sec:Conclusion}
We reviewed and discussed  the chiral matter problem, constituting 5\,\% in the cosmic energy pie, from the GUT point of view. In particular,
\begin{itemize}
\item
It is shown that anti-SU($N$) is a possibility. In this talk, anti meant that the GUT group breaking is achieved by the anti-symmetric tensor field(s) $\langle \Phi^{[AB]}\rangle= \langle\Phi_{[AB]}\rangle$. 

\item We argued for an anti-SU(7) $( = SU(7) \times U(1))$ families-unification model from a $\Z_{12-I}$ orbifold compactification. The $\Z_{12-I}$ orbifold is briefly discussed and the multiplicity 2 in the $T_3$ twisted sector is the key obtaining three chiral families. 

\item
Yukawa coupling structure is shown to be promising. A numerical study shows that the  anti-SU($7$) model satisfies the CKM fit \cite{KimCKMGUT}. 

\item It is shown that the doublet-triplet splitting is obtained naturally. Related to this, we commented that the dominant process for proton decay is by the exchange of GUT scale gauge bosons, implying $p\to \pi^0+e^+$ is the channel to search for. 

\item We also commented that  $\Z_{8-II}$ and  $\Z_{12-II}$ orbifolds are worthwhile to look for, and especially  SO(32) heterotic string can be useful for phenomenology.

\end{itemize}

 Note added after the talk: The anti-SU(7) model predicts $\dell=-\delq$ if only one GUT scale singlet $X$ develops a complex VEV which is the sole source of the weak CP violation \cite{KimCKMPMNS}.
 

\
\end{document}